\documentclass[12pt]{article}

\usepackage{color}

\newcommand{\be}{\begin{equation}}
\newcommand{\ee}{\end{equation}}
\newcommand{\ba}{\begin{eqnarray}}
\newcommand{\ea}{\end{eqnarray}}
\newcommand{\bea}{\begin{array}}
\newcommand{\eea}{\end{array}}

 \newcommand{\CB}{{\cal B}}
\newcommand{\CQ}{{\cal Q}}  \newcommand{\CT}{{\cal T}}
 
\newcommand{\CL}{{\cal L}} \newcommand{\CE}{{\cal E}}
\newcommand{\CN}{{\cal N}}

\newcommand{\CF}{{\cal F}}

\newcommand{\CP}{{\cal P}} 
\newcommand{\tr}{{\rm tr}}

\begin{document}

\begin{titlepage}

\begin{center}

\hfill  KIAS-P06009 \\
\hfill hep-th/0603179

\vspace{2cm}

{\bf \large Supertubes in Field Theories}

\vspace{1.4cm}

{  Seok Kim, Ki-Myeong Lee, Ho-Ung Yee}

\vspace{0.5cm}

{\it Korea Institute for Advanced Study, Seoul 130-722, Korea}


{\small e-mail: seok@kias.re.kr, klee@kias.re.kr,
ho-ung.yee@kias.re.kr}

\end{center}

\vspace{1.5cm}

\begin{abstract}
To a  domain wall or string object, Noether charge and topological
spatial objects can be attracted, forming a composite BPS
(Bogomolny-Prasad-Sommerfield) object. We consider two field
theories and derive a new BPS bound on composite linear solitons
involving multiple charges. Among the BPS objects `supertubes'
appear when the wall or string tension is canceled by the bound
energy, and could take  an arbitrary closed curve. In our theories,
supertubes manifest as Chern-Simons solitons, dyonic instantons,
charged semi-local vortices, and dyonic instantons on vortex flux
sheet.
\end{abstract}

\end{titlepage}

\baselineskip 18pt

In last few years there has been some interest in supertubes, which
are BPS objects of tubular shape with cross section of arbitrary
shape. Initially supertubes as the bound state of D2, D0 branes and
fundamental strings (F1) have been found by studying the
Dirac-Born-Infeld (DBI) action of D2 branes\cite{Mateos:2001qs}.
Later they have been found in many other
context\cite{Bak:2001kq,Emparan:2001ux,Bak:2001xx,Mateos:2001pi}.
The DBI description of static D2 branes of tubular shape with
uniform magnetic flux  and electric flux along tube direction shows
that the D2 brane tension could be canceled by the bound energy of
flux and charge, leading to the energy to be the sum of those for D0
and F1 branes. There would be a linear momentum of constant
magnitude flows  around the tube, which could lead to nonzero
angular momentum.

In this work we  ask whether such description suggested by  the DBI
action analysis of D2 brane is possible in  field theories. In a
couple of the theories considered here, we find that straight linear
topological structures  attract both Noether charges and topological
spatial solitons and form composite linear BPS objects, with a
linear momentum along the line. Among them, there are supertubes, in
which the bound energy cancels the tension of the linear structure.
Supertubes remain BPS when the linear structure gets bended. In
these theories, supertubes manifest as Chern-Simons solitons, dyonic
instantons, charged semi-local strings, and dyonic instantons on
vortex sheet at Higgs phase, and some more complicated composite
configurations.

In field theory, composite BPS objects usually involve only two
charges. BPS dyons typify one kind with the energy
$\sqrt{\CQ^2+\CT^2}$, where $\CQ$ and $\CT$ are the energies for two
charges. Clearly the charge is attractive to magnetic monopoles.
Another type is typified by q-lumps whose BPS energy is
$|\CQ|+|\CF|$ where $\CQ$ is the energy for the Noether charge and
$\CF$ is that for the lump\cite{Leese:1991hr}. Supertubes in field
theories are the generalization of q-lumps, and in addition, the
interior of supertubes have closed loops of domain wall or string in
arbitrary shape. Dyonic instantons and generalizations have been
studied as field theoretic
supertubes~\cite{Kim:2003gj,Townsend:2004nc}.

The composite linear BPS objects may have the energy contributions
from five sources. The original linear BPS structure has the tension
$\CT$. Noether charge leads to the energy density $\CQ$, and the
spatial topological charge does to $\CF$. There is an induced linear
momentum density $\CP$ along the line, which is not quite
independent of $\CQ$ and $\CF$. In addition the linear structure may
be imbedded in additional structure which cost energy density
$\CE_0$. We find the BPS energy per unit length to be
\be \CE_{line} = \CE_0+\sqrt{(\CF\pm \CQ)^2+(\CT\mp\CP)^2}.
\label{1bps} \ee
This shows what the energy of the composite BPS object is made of.
For the supertube case, $|\CT|=|\CP|$,  the BPS energy density
becomes $\CE_{tube}=\CE_0 +|\CF|+|\CQ|$, indicating that the tension
of the linear structure does not matter anymore. All cases the
diagonal component of the energy momentum tension along the momentum
direction vanishes, which is another indication that it does not
cost any energy to bend the linear structure.  The linear momentum
density $\CP$ is given by the original tension $\CT$, and so have
constant magnitude. For supertubes of linear shape, one can have
$\CF$ and $\CQ$ arbitrary such that $\CF\CQ$ is fixed. (If
$|\CF|=|\CQ|$, the BPS configurations are chiral waves along the
linear structure.)

Let us start with an abelian Chern-Simons Higgs
theory~\cite{Hong:1990yh,Jackiw:1990pr,Jackiw:1990aw}, whose
Lagrangian  is
 \be \CL = \frac{\kappa}{2}
\epsilon^{\mu\nu\rho}A_\mu\partial_\nu A_\rho - \frac{1}{2}|D_\mu
\phi |^2- \frac{1}{8\kappa^2}|\phi|^2(|\phi|^2-v^2)^2 , \ee
where $\epsilon^{012}=1$  and  $D_\mu \phi = (\partial_\mu +iA_\mu)
\phi$. There are two degenerated vacua, the symmetric phase $\phi=0$
and the asymmetric phase $\phi=v$, which allow BPS domain walls.
There are Chern-Simons solitons in both symmetric and broken phases.
The Gauss law is $ \kappa F_{12} +\frac{i}{2}(D_0\phi^*\phi-
\phi^*D_0 \phi ) = 0 $. The Chern-Simons solitons are
charge-magnetic flux composite objects, carrying fractional spin and
satisfying fractional statistics.

For any field configuration, we rewrite  the energy per unit length
along a chosen direction, say $x^2$,  as
\ba \CE \!\! &=&\!\! \frac{1}{2}\int dx^1\! \left\{ \biggl|D_0\phi -
\frac{i}{2\kappa}\phi
(|\phi|^2-v^2)\cos\alpha + D_2\phi\sin\alpha \biggr|^2 \right. \nonumber \\
& & \left. + \biggl|D_1\phi -i D_2\phi
\cos\alpha +\frac{1}{2\kappa}\phi(|\phi|^2-v^2)\sin\alpha\biggr|^2 \right\} \nonumber \\
& & + \CB \cos\alpha +(\CT- \CP) \sin\alpha\, , \label{cshe}\ea
where the magnetic flux energy density per unit length, the domain
wall energy density, and  the linear momentum along the domain wall
are, respectively,
\ba && \CB = \frac{v^2}{2}\int dx^1 F_{12}, \;\; \CT =
\frac{1}{8\kappa} \int dx^1 \;
\partial_1( 2v^2 |\phi|^2- |\phi|^4),\nonumber \\
&&  \CP= \frac{1}{2} \int dx^1 ( D_0\phi^*
D_2\phi+D_2\phi^*D_0\phi). \ea
As all terms in the integration of Eq.(\ref{cshe}) is nonnegetive
for any angle $\alpha$, we have the bound on the energy density
\be \CE\ge \, \CE_{line}\equiv \sqrt{\CB^2 +(\CT - \CP)^2} \ge \,
\CE_{tube}\equiv|\CB|\, . \ee
Note that in the Chern-Simons theory, the Gauss law relates the
Noether charge with topological magnetic flux. The composite of
domain wall and solitons have been studied before but without the
above result~\cite{Kao:1996tv}.

Among this linear BPS configurations which saturate the bound
$\CE=\CE_{line}$, the special ones are those satisfy $ \CT =\CP$, so
that $\cos\alpha= \pm 1$. The energy would be determined purely by
the magnetic flux as $ \CE_{tube}=|\CB|$. Thus the wall energy is
canceled by the bound energy of the wall and charge, and so the
domain wall can bend. This supertube configurations satisfy the
Gauss law and the self-dual Chern-Simons soliton equation,
\be D_0\phi \mp \frac{i}{2}\phi(|\phi|^2-v^2)=0,\;\; D_1\phi\mp
iD_2\phi=0 ,\label{cssd}\ee
which is the equation studied well in
Ref.\cite{Hong:1990yh,Jackiw:1990aw,Jackiw:1990pr}. Thus, if we
imagine the effective string action for the supertube, it cost no
energy to bend  the domain wall. To find explicitly  the linear
configuration of the supertube, we choose $\cos\alpha=1$ and
$\kappa>0$ with ansatz $\phi=|\phi|(x^1)e^{-iv^2(x^0+x^2)/2\kappa}$,
$A_2(x^1)=A_0(x^1)$. The self-dual equations (\ref{cssd}) and the
Gauss law  can be solved with noting $A_0= |\phi|^2/2\kappa$. The
solution is given by  $|\phi|^2(x^1) = v^2 e^{v^2x^1/2\kappa}(1+
e^{v^2x^1/2\kappa})$ with the boundary condition $\phi=0 $ at
$x=-\infty$ and $|\phi|=v$ at $x=\infty$. From this we can find
$\CT=\CP= v^4/(8\kappa)$ and $\CF= v^4/(4\kappa)$. Note that the
direction of the momentum flow is fixed by the orientation of the
domain walls. This is the case for the Chern-Simons solitons.

While many properties of Chern-Simons solitons are known, we obtain
somewhat new perspective of these objects from the fact that they
can be regarded as supertubes. (The tubular direction is unclear in
this theory at the moment.) In large magnetic flux limit, vortices
can be regarded as a collection of supertubes inside of which is in
the symmetric phase and outside of which is in the asymmetric phase.
Nontopological solitons in large charge limit can be regarded as a
collection of supertubes inside of which is in the asymmetric phase
and outside of which is in the symmetric phase. One can have
nontopological solitons with vortices, which can be regarded as
supertubes in supertubes. Almost linear structure of the large flux
limit suggests more. For example, the maximal angular momentum can
be achieved by a single supertube of circular shape. Indeed, there
is such a bound on angular momentum on Chern-Simons
solitons~\cite{Jackiw:1990pr}.  The supersymmetry of these solitons
has been studied in $\CN=2$ Chern-Simons theory and shown to be 1/2.


The second theory we consider here is a $\CN=1$ U(N) gauge theory
with $N_f=0$ or $N_f\ge N$ flavors in 5-dim spacetime. For
simplicity, we consider the bosonic fields, which are made of the
gauge multiplet $A_M, A_5=\phi$ and the flavor multiplets  $q_f,
q'_f, f=1..N_f$. Here $q'_f$ vanish for the BPS configurations and
so are neglected. The bosonic part of the Lagrangian is
\ba {\cal L} \!\! &=&\!\!
\frac{1}{2e^2}\tr\biggl(-\frac{1}{2}F_{MN}F^{MN}+D_M\phi
D^M\phi - D^2\biggr) \nonumber \\
& & \!\! + \frac{1}{2}\tr\bigg( D_M\bar{q}_f D^M q_f -(\phi-m_f)^2
\bar{q}_f q_f \biggr), \ea
where the nonvanishing D-term is $D=D^3 = e^2(v^2-\bar{q}_f q_f)/2$
and $D_Mq_f = \partial_Mq_f+ iq_f A_M$. The Gauss law is
\be D_\mu F_{\mu 0} +i[\phi,D_0\phi] +\frac{ie^2}{2} (\bar{q}_f
D_0q_i-D_0\bar{q}_f q_f)=0. \ee

The supersymmetry transformation of the guagino field $\lambda_i$
and the matter fermion $\psi_f$ is given as (see, for
example~\cite{Lee:2005sv,Eto:2005sw} .)
\be \!\! \delta \lambda_i = \frac{1}{2}F_{MN}\Gamma^{MN}\epsilon_i +
i{\bf D}^a\sigma^a_{ij}\epsilon_j ,\;   \delta\psi_f =
D_Mq_{fi}\Gamma^M\epsilon_i\,   \ee
in 6-dim notation with symplectic Majorana-Weyl spinor $\epsilon_i,
\lambda_i$. For our theory, $q_{f1}=q_f, q_{f2}=q'_f$ and
$D_5q_f=q_f(\phi-m_f)$.  We impose an 1/4 BPS condition on the
spinor parameter by imposing two conditions,
\be
\Gamma^{34}e^{\alpha\Gamma^{45}}i\sigma^3_{ij}\epsilon_j=\epsilon_i,
\;\; \Gamma^{05}e^{\alpha\Gamma^{45}}\epsilon_i=\pm \epsilon_i ,\ee
which implies $\Gamma^{12}i\sigma^3_{ij}\epsilon_j=\mp\epsilon_i $
as $\epsilon_i$ is chiral. For such constant nonzero spinor
$\epsilon_i$, the supersymmetric transformations $\delta \lambda_i$
and $\delta \psi_f$ vanish if the bosonic configurations satisfy the
BPS equations
\ba && B_i \pm \delta_{i3} D^3 \mp F_{i4}\cos\alpha \mp
D_i\phi\sin\alpha=0,\;\nonumber\\
&& F_{i0} \pm F_{i4}\sin\alpha\mp  D_i \phi\cos\alpha = 0, \nonumber\\
&& F_{40}\mp D_4\phi\cos\alpha=0, D_0\phi\pm D_4\phi\sin\phi=0, \nonumber \\
&&  (D_1\pm i D_2)q_f=0, \nonumber \\
&&  D_0q_f\pm D_4q_f \sin\alpha\mp iq_f(\phi-m_f)\cos\alpha=0, \nonumber \\
&&  D_3q_f-iD_4q_f \cos\alpha +q_f(\phi-m_f)\sin\alpha=0, \ea
where $B_3=F_{12}$ and so on.  The energy density is given by the
square of the above terms with proper coefficient and the boundary
terms after the Gauss law is used. As the square terms are positive
definite,  we get a bound on the energy density
\be \CE \ge \mp\frac{v^2}{2}\tr F_{12} + (\CF\pm \CQ)\cos\alpha
+(\CT\mp \CP)\sin\alpha \ee
where $\alpha$ is arbitrary and
\ba
&&\!\!\!\!\!\!\!\!\!\CF= \frac{v^2}{2}\tr F_{34} \pm\frac{1}{e^2}\tr B_i F_{i4},\nonumber \\
\!\!\!\!\! && \!\!\!\!\!\!\!\!\!\CQ = \frac{1}{e^2}\partial_\mu \tr
(F_{\mu 0}\phi) +\frac{i}{2} m_f \tr(\bar{q}_f D_0 q_f
-D_0\bar{q}_f q_f), \; \nonumber \\
&&\!\!\!\!\!\!\!\!\! \CT= \pm \frac{1}{e^2}\partial_i \tr(B_i\phi)
+\frac{1}{2}\partial_3\tr\biggl(v^2\phi -\bar{q}_f q_f
(\phi-m_f)\biggr), \nonumber \\
&&\!\!\!\!\!\!\!\!\!\CP = \frac{1}{e^2}\tr(F_{i0}F_{i4}+D_0\phi
D_4\phi) \nonumber \\
&& +\frac{1}{2} \tr(D_0\bar{q}_fD_4q_f +D_4\bar{q}_f D_0q_f) . \ea
The first term $F_{12}$ term comes from the  magnetic flux vortex
sheet. The linear structure of magnetic monopole string or domain
wall lead to the tension part $\CT$.  The linear momentum density
along $x^4$ is $\CP$. $\CQ$ is the energy due to the flavor charge.
$\CF$ is the energy density due to the magnetic flux $\tr F_{34}$
and the instanton density. We have to integrate over the transverse
direction to the linear structure to get the energy density, leading
to the BPS energy density (\ref{1bps}).  In a gauge $A_0=\pm
\eta\phi$, the supertube configurations, which saturate the energy
bound with $\cos\alpha=\eta=\pm 1$, satisfy the following equations,
\ba \!\!\!\! && \!\!\!\!\! B_i\mp \eta F_{i4}\pm
\delta_{i3}\frac{e^2}{2}(v^2-\bar{q}_f q_f) = 0\, ,
 \nonumber \\
\!\!\!\! && \!\!\!\!\! (D_1\pm iD_2)q_f=0,\, (D_3 -i\eta D_4) q_f=0\, , \nonumber \\
\!\!\!\! && \!\!\!\!\! D_i^2 \phi
-\frac{e^2}{2}\biggl((\phi-m_f)\bar{q}_fq_f
+\bar{q}_fq_f(\phi-m_f)\biggr)=0\, . \label{2bps}\ea
Note that the first three equations are for the spatial structure.
The last one is additional structure on the spatial structure.
Typically the spatial solitons have some scaling parameters, which
collapse when some symmetry is further broken. But the repulsive
force among Noether charge balances the collapsing force, making
supertubes as  dyonic spatial solitons.

We will consider the four supertube examples in this theory. First
of all, let us neglect all flavor. The supersymmetry can enlarged to
$\CN=2$ in 5-dim if we add an adjoint matter multiplet. Assume the
gauge group is $SU(2)$. Then the obvious linear object is the
monopole string, to which electric charge and instantons are
attracted. The integration all above densities over three
dimensional space transverse monopole string leads to the BPS bound
(\ref{1bps}) on the energy density along the string.  (Here we are a
bit loose on the notation for the energy densities and its
integrated quantities.) Here the 16 susy leads to more sign freedom
and so more sign combinations are allowed. When $\cos\alpha=\pm 1$,
we get the supertube configuration, whose BPS equation (\ref{2bps})
becomes that for dyonic instantons, in which arbitary shaped closed
loops of magnetic monopoles are present.

For a single monopole string with known $A_i, \phi$, the linear
density for these quantities are trivial to get with $\phi=vA_4/h$
where $A_4= h\sigma^3/2$ and $\phi=v\sigma^3/2$ asymptotically. They
are $\CT=\CP=4\pi v/e^2$ and $\CF=4\pi h/e^2$ and $\CQ=4\pi v^2/e^2
h$, and so $\CQ\CF$ is independent of $h$. This tell us the
characterstics of dyonic instantons in large instanton and charge
limit, where the linear approximation for the monopole strings may
work. Dyonic instantons have been studied studied in many
directions~\cite{Lambert:1999ua,Eyras:2000dg}. The monopole string
composite with charge and instantons have been also
studied~\cite{Bak:2002ke,Lee:2002xh}. Monopole strings inside dyonic
instantons appear when the instanton number is at least two, showing
they are supertubes\cite{Kim:2003gj}.

Second case is the $U(1)$ theory with $N_f=2$. When the flavor
symmetry is nonabelian with $m_1=m_2$, the theory allows semi-local
vortices~\cite{Vachaspati:1991dz,Hindmarsh:1991jq}.  With $m_1<m_2$,
there are two degenerated vacua with $q_1=v,q_2=0,\phi=m_1$ and
$q_1=0,q_2=v,\phi=m_2$, which allows a domain
wall~\cite{Tong:2002hi,Isozumi:2003rp}. With $m_1<m_2$, the core of
the semi-local vortices collapse.  We ignore $x^1$ and $x^2$
direction completely. The domain wall attracts the flavor charge and
magnetic flux $F_{34}$. For energy density along the line $x^4$, we
integrate over $x^3$, with $\CF$ due to the magnetic flux $F_{34}$.
Again we obtain BPS energy bound (\ref{1bps}) per length. Especially
for $\cos\alpha=\pm 1$, the domain wall can bend, and the BPS
equations (\ref{2bps}) becomes those for charged semi-local
vortices~\cite{Abraham:1992hv}. For a straight domain wall with
ansatz $q_f=|q_f|e^{-m_fx^0-ik_f x^3}$ with field choice so that
$m_2=-m_1=m>0$ and $k_2=-k_1=k>0$, with appropriate boundary
condition, we see that $A_4=k\phi/m$ and so obtain that
$\CT=\CP=mv^2$ and $\CF=kv^2$ and $\CQ=mv^2/2k$. Again $\CF\CQ$ is
independent of $k$ and the component $T_{44}$ along the domain wall
direction vanishes.

A supertube for generic charged semi-local  vortices  of $U(1)$
theory with two flavors of different mass with a large magnetic flux
can be caricatured as follows. Say the vacuum is where $q_1=v,
q_2=0$. The vortex lies in 3-4 plane, and we ignore 1,2 direction.
For large vorticity, there is an arbitrary shaped closed line of
domain wall,  outside of which $|q_1|=v$ and $q_2=0$, and inside of
which $q_1=0$ and $|q_2|=v$. Along the line there are energy density
which is identical to the momentum density given above. There are
nontrivial magnetic flux and flavor charge which can vary along the
line as long as their product is fixed as in the previous paragraph.

The third case appears in the  $U(2)$ gauge theory with two flavors.
The vacuum is in the color-flavor locking phase $\bar{q}_1=(v,0),
\bar{q}_2=(0,v)$ and $\phi={\rm diga}(m_1,m_2)$. We consider a
vortex sheet on the 3-4 plane at $x^1,x^2\sim 0$ with nonzero
$F_{12}$, and with a magnetic monopole string along $x^4$ direction
at $x^3\sim 0$ when $m_1\ne m_2$~\cite{Tong:2003pz,Shifman:2004dr}.
The gauge orientation of the magnetic flux $F_{12}$ changes from
${\rm diag}(1,0)$ to ${\rm diag}(0,1)$ as $x^3$ goes from $-\infty$
to $\infty$.  Thus the energy would have a contribution $\CE_0$ from
the magnetic flux $\tr F_{12}$ which is localized near $x^1=x^2=0$
and the monopole string tension $\CT$. Monopole strings attract
flavor charge and instantons, which leads to energy contribution
$\CQ$ and $\CF$. Integration over $x^1,x^2,x^3$ with some
appropriate infrared cutoff in $x^3$ direction would lead to a
energy density bound (\ref{1bps}). The BPS equations (\ref{2bps})
becomes that for dyonic instantons on the vortex sheet, where
$q_f\sim e^{-im_f x^0}$. When $m_1=m_2$, one can have instantons in
the vortex sheet in the Higgs phase\cite{Eto:2004rz}, which
collapses $m_1\ne m_2$. Flavor Noether charges keep them from
collapsing.

For the linear configuration with additional ansatz $q_f \sim
e^{-ik_f x^4}$, and we shift the fields $A_4$ and $\phi$ so that
$m_1=-m_2=m/2$ and $k_1=-k_2=k/2$, which implies asymptotic value of
$A_4={\rm diag}(k,-k)/2$, and $\phi={\rm diag}(m,-m)/2$. The
selfdual equation and the gauss law are compatible if $A_4=k\phi/m$.
For a single monopole string, we obtain $\CT=\CP= 4\pi m/e^2$ and
$\CF=4\pi k/e^2$ and $\CQ= 4\pi m/k$ after some effort. Again
$\CF\CQ$ is independent of $k$.

Then for large instanton number and flavor charge, the supertube
configuration can be pictured as follows. We have a flat vortex
sheet on which an arbitrary shaped closed curve of monopole string
is present. On the sheet, outside of the curve $q_1=0, q_2={\rm
diag}(0,v)$ at the sheet center and  inside of the curve $q_1={\rm
diag}(v,0)$ and $q_2=0$. On the curve, there are nonzero flavor
charge and instanton density per unit length, their product remain
uniform. (It is of course not well defined unless the curve can be
approximated to be straight.) Along the curve, there is a linear
momentum flows with unform magnitude given by the monopole tension.
This configuration would become  dyonic instanton in the Coulomb
when the FI parameter $v^2$ approaches zero.

For the last case, we consider the theory with $U(2)$ gauge group
and $N_f=3$ case. As there are degenerated vacua, one can have
domain walls. In addition, one can have vortices sheet with magnetic
monopoles strings, which allows a composition of
vortex-monopole-domain wall~\cite{Sakai:2005sp,Auzzi:2005yw}. Our
BPS equation allows adding instanton numbers, flavor charges and
additional flux $F_{34}$. Consequently we can imagine a composition
of  charged semi-local vortices and an dyonic instanton sin the
vortex sheet. The interpretation of the above field theory as the
field theory on D4-D8 branes seems to show above supertubes in nice
figurative way.

To conclude, we found a new BPS energy bound on straight composite
linear BPS objects. Especially supertubes seem to be everywhere when
one can add Noether charge to topological solitons. While our
analysis has been done on relatively simpler systems, one can
generalization to nonabelian gauge groups and more matter fields,
which would allow richer supertube structures.

This work is supported in part  by KOSEF Grant R01-2003-000-10319-0,
KOSEF SRC Program through CQUeST at Sogang (K.L.), and  Korean
Research Foundation Grant No. KRF-2005-070-C00030 (K.L.).  The
authors (K.L. and H.U.Y.) thank the Yukawa Institute for Theoretical
Physics at Kyoto University, where the part of the work was
motivated  during the YITP-W-05-21 (a workshop) on ``Fundamental
Problems and Applications of Quantum Field Theory''.

\end{document}